\documentclass[12pt]{article}

\usepackage{amssymb}
\usepackage{amsmath}
\usepackage{latexsym}
\usepackage{yfonts}

\catcode`@=11
\renewcommand{\section}{\@startsection{section}{1}{0pt}{\medskipamount}
{\medskipamount}{\large\bf}} \numberwithin{equation}{section}
\catcode`@=12
\def\beq{\begin{eqnarray}}    %%%  begequation/eqnarray
\def\eeq{\end{eqnarray}}      %%%  endequation/eqnarray

\def\ln{\,\mbox{ln}\,}                  %%% log
\def\Box{\square}                       %%% Box
\def\pa{\partial}                       %%% partial
\def\={\ =\ }

\begin{document}

\begin{center}

{\Large\bf Gauge dependence of  effective average action}

\vspace{18mm}

{\large
Peter M. Lavrov$\footnote{E-mail:
lavrov@tspu.edu.ru}$\; }

\vspace{8mm}

{\em
Tomsk State Pedagogical University,\\
Kievskaya St.\ 60, 634061 Tomsk, Russia}

\vspace{20mm}

\begin{abstract}
\noindent
The gauge dependence of effective average action
in the functional renormalization group is studied.
The effective average action is considered
as non-perturbative solution to
the flow equation which is the basic equation of the method.
It is proven that at any scale of IR cutoff the effective average action depends
on gauges making impossible physical interpretation of all obtained results
in this method.

\end{abstract}

\end{center}

\vfill

\\

\newpage

\section{Introduction}
\noindent It is a well-known fact that Green functions in gauge
theories (and therefore the effective action being the generating
functional of one-particle irreducible Green function or vertices)
depend on gauges. On the other, hand elements of S-matrix should be
gauge independent. It means that the gauge dependence of effective
action should be a very special form. The gauge dependence is a
problem in quantum description of gauge theories beginning with
famous papers by Jackiw \cite{Jac} and Nielsen \cite{Niel} where the
gauge dependence of effective potential in Yang-Mills theories has
been found. For Yang-Mills theories in the framework of the
Faddeev-Popov quantization method \cite{FP} the gauge dependence
problem has been found in our papers \cite{LT1,LT2} and  for general
gauge theories within the Batalin-Vilkovisky formalism \cite{BV,BV1}
in our paper \cite{VLT} respectively.

Over the past three decades, there has been an increased interest in the
nonperturbative approach in Quantum Field Theory known
as the functional renormalization group (FRG)  proposed
by Wetterich \cite{Wet1,Wet2}. The FRG approach  has got further
developments
and numerous applications. There are many reviews  devoted
to detailed discussions of different aspects of the FRG approach
and among them one can find \cite{Paw,Rost,Gies}
with qualitative references.
As a quantization procedure the FRG  belongs to covariant quantization schemes
which meets in the case of gauge theories  with two principal problems: the
unitarity of S-matrix and the gauge dependence of results obtained.
Solution to the unitarity problem requires
consideration of canonical formulation of a given theory
on quantum level and use of the Kugo-Ojima method \cite{KO}
in construction of physical state  space. Within the FRG the
unitarity problem is not considered at all because main efforts are connected
with finding solutions to the flow equation for the effective average action.

In turn the gauge dependence problem exists for the FRG approach as unsolved ones
if one does not take into account the reformulation based
on composite operators \cite{LSh}
where the problem was discussed from point of view the
basic principles of the quantum field theory.
Later on the gauge dependence problem in the FRG was discussed
in our papers several times for Yang-Mills and quantum gravity theories
\cite{L,LRNSh,Lav2,BLRNSh,Lav1}
 but the reaction from the FRG community was very weak
 and came down only to mention without any serious study.
Situation with the gauge dependence in the FRG
 is very serious because without solving the problem
 a physical interpretation of results obtained is impossibly.
 It is main reason to return for discussions of the gauge
 dependence problem of effective average action in the FRG approach.

The DeWitt's condensed notations \cite{DeWitt} is used.
The functional derivatives with respect to
fields and sources are understood  as right and left ones respectively.

\section{Gauge dependence in Yang-Mills theories}

We start with the action $S_0[A]$ of fields $A$ for given Yang-Mills theory.
 Generating functional of Green functions, $Z[J]$,
 can be constructed by the Faddeev-Popov rules \cite{FP}
 in the form of functional integral
\beq
\label{n1}
Z[J]=\int D\phi
\;\exp\Big\{\frac{i}{\hbar}\big(S_{PF}[\phi]
+J_i\phi^i \big)\Big\}.
\eeq
where $\phi=\{\phi^i\}=(A,B,C,{\bar C})$ is full set of fields including the ghost $C$ and
antighost ${\bar C}$ Faddeev-Popov fields and auxiliary fields $B$
(Nakanishi-Lautrop fields), $J=\{J_i\}$ are external sources to fields
$\phi$, $S_{PF}[\phi]$ is the Faddeev-Popov action
\beq
\label{n2}
S_{PF}[\phi]=S_0[A]+\Psi[\phi]_{,i}R^i(\phi).
\eeq
Here in (\ref{n2}) $\Psi[\phi]$ is gauge fixing functional (in the simplest case having the
form $C\partial A$),
%$R^i(\phi)$ are generators of BRST transformations %($\delta_B \phi^i=R^i(\phi)\mu, \;\mu={\rm const},\;\mu^2=0$)
 and notation
$X_{,i}=\delta X/\delta \phi^i$ is used.
The Faddeev-Popov action $S_{FP}[\phi]$ (\ref{n2}) obeys very important property of
invariance under global supersymmetry - BRST
symmetry \cite{brs1,t},
\beq
\label{n3}
\delta_B S_{FP}[\phi] =0,\quad  \delta_B\phi ^i=R^i(\phi)\mu,
\quad \mu^2=0,
\eeq
where $R^i(\phi)$ are generators of BRST transformations. In what follows the explicit
form of $R^i(\phi)$ is not essential and we omit their presentation.

From definitions (\ref{n1}) and (\ref{n2}) it follows that the functional $Z[J]$ depends oh gauges
To study the character of
this dependence, let us consider an infinitesimal variation of gauge fixing
functional $\Psi[\phi]\;\rightarrow\;\Psi[\phi]+\delta\Psi[\phi]$ in
the functional integral for $Z[J]$. Then we obtain ($\pa_{\!J}=\delta/\delta J$)
\beq
\nonumber
\delta Z[J]&=&\frac{i}{\hbar}\int D\phi\; \delta\Psi_{,i}[\phi]R^i(\phi)
\;\exp\Big\{\frac{i}{\hbar}\big(S_{PF}[\phi]
+J_i\phi^i\big)\Big\}=\\
\label{n4}
&=&\frac{i}{\hbar}
\delta\Psi_{,i}[-i\hbar\pa_{\!J}]
R^i(-i\hbar \pa_{\!J})Z[J].
\eeq

There exists an equivalent presentation of the variation for $Z[J]$ under
variations of gauge conditions. Indeed,
making use the change of integration variables in the functional integral
for Z[J] with the  choice $\Psi[\phi]+\delta \Psi[\phi]$ in the form of the
BRST transformations,
\beq
\label{n5}
\delta\phi^i=R^i(\phi)\mu[\phi],
\eeq
taking into account that  the corresponding Jacobian, $J$,
is equal to
\beq
\label{n6}
J=\exp\{-\mu[\phi]_{,i}R^i(\phi)\},
\eeq
choosing the functional $\mu[\phi]$ in the form
$\mu[\phi]=(i/\hbar)\delta\Psi[\phi]$, then we have
\beq
\nonumber
\delta Z[J]&=&\frac{i}{\hbar}\int D\phi\;
J_iR^i(\phi)\delta\Psi[\phi]\exp\Big\{\frac{i}{\hbar}\big(S_{PF}[\phi]
+J_i\phi^i\big)\Big\}=\\
\label{n7}
&=&
\frac{i}{\hbar}J_iR^i(-i\hbar\pa_{\!J})\;
\delta\Psi[-i\hbar\pa_{\!J}]\;Z[J].
\eeq
Both relations  are equivalent
due to the evident equality
\beq
\label{n8}
\int D\phi \;\pa_{\!\phi^j}\Big(\Psi[\phi]R^j(\phi)
\exp\Big\{\frac{i}{\hbar}\big(S_{PF}[\phi]
+J_i\phi^i\big)\Big\}\Big)=0,
\eeq
where the following equations
\beq
\label{n9}
S_{PF,i}[\phi]R^i(\phi)=0,\quad R^i_{\;,i}(\phi)=0,\quad
R^i_{\;,j}(\phi)R^j(\phi)=0,
\eeq
should be used.

In terms of the functional $W[J]=-i\hbar\ln Z[J]$ the above relations
 rewrite as
\beq
\label{n10}
&&\delta W[J]=J_i
R^i(\pa_{\!J}W-i\hbar \pa_{\!J})\;
\delta\Psi[\pa_{\!\!J}W-i\hbar \pa_{\!J}]
\cdot 1,\\
\label{n11}
&&\delta W[J]=
\delta\Psi_{,i}[\pa_{\!J}W-i\hbar\pa_{\!J}]
R^i(\pa_{\!J}W-i\hbar\pa_{\!J})\cdot 1.
\eeq

Introducing the effective action, $\Gamma=\Gamma[\Phi]$,
through the Legendre transformation of $W[J]$,
\beq
\label{n12}
\Gamma[\Phi]=W[J]-J_i\Phi^i, \quad \Phi^i=\frac{\delta W}{\delta J_i},
\quad \frac{\delta \Gamma}{\delta\Phi^i}=-J_i,
\eeq
the gauge dependence of effective action
 is described by the equivalent relations
\beq
\label{n13}
&&\delta\Gamma[\Phi]=-\frac{\delta \Gamma}{\delta\Phi^i}\;
R^i({\hat \Phi})\;\delta\Psi[{\hat \Phi}]\cdot 1,\\
\label{n14}
&&
\delta\Gamma[\Phi]=
\delta\Psi_{,i}[{\hat \Phi}]\;
R^i({\hat \Phi})\cdot 1,
\eeq
where the notations
\beq
\label{n15}
{\hat \Phi}^i=\Phi^i+ i\hbar(\Gamma^{''-1})^{ij}\,
\frac{\delta}{\delta\Phi^j},\quad
(\Gamma^{''})_{ij}=\frac{\delta^2\Gamma}{\delta \Phi_i\delta\Phi_j},\quad
\big(\Gamma^{''-1}\big)^{ik}\cdot
\big(\Gamma^{''}\big)_{kj}\,=\delta^i_{\,j},
\eeq
are used.

From the presentation (\ref{n13}) it follows the important statement
that the effective action does not depend on the gauge conditions at
the  extremals, \beq \label{n14}
\delta\Gamma\big|_{\pa_{\Phi}\Gamma=0}=0, \eeq making possible the
physical interpretation of results obtained in the Faddeev-Popov
-method for Yang-Mills theories.

There exists  another description of gauge dependence
of effective action for Yang-Mills theories:
The effective action can be presented in the form of gauge independent
functional in which all gauge dependence contains in   their arguments \cite{LT1,LT2}.

\section{Effective average action}

Main idea of the functional renormalization group
is to modify the behavior of propagators in IR region with
the help of a regulator action $S_k[\phi]$ being quadratic in fields.
In the case of Yang-Mills theories it leads to action
\beq
\label{m1}
S_{Wk}[\phi]=S_{FP}[\phi]+S_k[\phi],\quad
S_k[\phi]=\frac{1}{2}R_{k|ij}\phi^j\phi^i.
\eeq
Standard choice of regulators $R_{k|ij}$ is
\beq
\label{m2}
R_{k|ij}=z_{ij}\frac{\Box \exp\{-\Box/k^2\}}{1-\exp\{-\Box/k^2\}},
\quad \Box=\pa_{\mu}\pa^{\mu},
\eeq
with properties
\beq
\label{m3}
\lim_{k\rightarrow  0}R_{k|ij}=0.
\eeq
The action $S_{Wk}[\phi]$ is not invariant under the BRST transformations
\beq
\label{m4}
\delta_B S_{Wk}[\phi]=\delta_B S_k[\phi]=R_{k|ij}\phi^jR^i(\phi)\mu\neq 0.
\eeq

The generating functional of Green functions has the form
\beq
\label{m5}
\!Z_k[J]\!=\!\int\!\!\! D\phi \,\exp \Big\{\frac{i}{\hbar}
\big[S_{Wk}[\phi]\!
+\!J_j\phi^j\big]\Big\}\!=\!\exp \Big\{\frac{i}{\hbar}
W_{k}[J]\Big\},
\eeq
Variation of $\delta Z_k[J]$ under change of gauge condition can be presented
as
\beq
\label{m6}
\delta Z_k[J]=
\frac{i}{\hbar}\big(J_i-i\hbar R_{k|ij}\pa_{\!J_j}\big)R^i(-i\hbar\pa_{\!J})\;
\delta\Psi[-i\hbar\pa_{\!J}]\;Z[J].
\eeq
In terms of $W_{k}[J]$ we have
\beq
\label{m7}
\delta W_k[J]\!=\!\big(J_i\!+\!R_{k|ij}(\pa_{J_j}W_k-i\hbar \pa_{J_j})\big)
R^i(\pa_{\!J}W-i\hbar\pa_{\!J})\;
\delta\Psi[\pa_{\!\!J}W-i\hbar\pa_{\!J}]
\cdot 1,
\eeq

Introducing the effective average action, $\Gamma_k=\Gamma_k[\Phi]$,
being the main quantity in the FRG  through the Legendre transformation of $W_k[J]$,
\beq
\label{m8}
\Gamma_k[\Phi]=W_k[J]-J_i\Phi^i, \quad \Phi^i=\frac{\delta W_k}{\delta J_i},
\quad \frac{\delta \Gamma_k}{\delta\Phi^i}=-J_i,
\eeq
the gauge dependence of effective average action
 is described  as %by the  relation
\beq
\label{m9}
\delta \Gamma_k[\Phi]=-\Big(\frac{\delta \Gamma_k}{\delta\Phi^i}
-R_{k|ij}{\hat \Phi}^j\Big)\;
R^i({\hat \Phi})\;\delta\Psi[{\hat \Phi}]\cdot 1,
\eeq
with the notations
\beq
\label{m10}
{\hat \Phi}^i=\Phi^i+ i\hbar(\Gamma_k^{''-1})^{ij}\,
\frac{\delta}{\delta\Phi^j},\quad
(\Gamma_k^{''})_{ij}=\frac{\delta^2\Gamma_k}{\delta \Phi_i\delta\Phi_j},\quad
\big(\Gamma_k^{''-1}\big)^{il}\cdot
\big(\Gamma_k^{''}\big)_{lj}\,=\delta^i_{\,j}.
\eeq
Due to (\ref{m9}) the effective average action remains gauge dependent
even on their extremals
\beq
\label{m11}
\delta\Gamma_k[\Phi]\Big|_{\pa_{\Phi}\Gamma_k=0}\neq 0
\eeq
making impossible physical interpretation of results obtained in the FRG.

\section{Gauge dependence of  flow equation}

The above analysis of gauge dependence of effective average action
$\Gamma_k[\Phi]$ does not convince people from the FRG community because
it is based on theorems in Quantum Field Theory formulated within standard
perturbation approach while it is assumed that the flow equation for
$\Gamma_k[\Phi]$ in the FRG is considered non-perturbatively.

We are going to study the gauge dependence of effective average action
found as a solution to the flow equation. To do this
 we find first of all the partial derivative of $Z_k[J]$ with respect
to IR cutoff parameter $k$.
The result reads
\beq
\nonumber
\pa_kZ_k[J]&=&\frac{i}{\hbar}\int D\phi \,
\pa_kS_k[\phi]
\exp \Big\{\frac{i}{\hbar}
\big[S_{Wk}[\phi]
+J_A\phi^A\big]\Big\}\\
\label{k1}
&=&\frac{i}{\hbar}
\pa_kS_k[-i\hbar\pa_{\!J}]Z_k[J].
\eeq
In deriving this result, the existence of functional integral
is only used.

In terms
of generating functional of connected Green functions we have
\beq
\label{k2}
\pa_k W_k[J]=\pa_k S_k[\pa_{\!J} W_k
-i\hbar\pa_{\!J}]\cdot 1 .
\eeq
The basic equation (flow equation)  of the FRG approach has the form
\beq
\label{k3}
\pa_k \Gamma_k[\Phi]=\pa_k S_k[{\hat \Phi}]\cdot 1,
\eeq
where ${\hat \Phi}=\{{\hat \Phi}^i\}$ is defined in (\ref{m10}).
It is assumed that solutions to the flow
equations  present the effective average action
$\Gamma_k[\Phi]$
beyond the usual perturbation calculations.

Now, we analyze the gauge dependence problem of
the flow equation . Note that up to now this problem has never been discussed
in the literature.
To do this we consider the variation of $\pa_kZ_k[J]$
under an infinitesimal change
of gauge fixing functional, $\Psi[\phi]\rightarrow \Psi[\phi]+\delta\Psi[\phi]$.
Taking into account that $\pa_kS_k[\phi]$ does not depend on
 gauge fixing procedure, we obtain
\beq
\label{k4}
\delta\pa_kZ_k[J]=\Big(\frac{i}{\hbar}\Big)^2
\pa_kS_k[-i\hbar\pa_{\!J}]
\delta\Psi_{,i}[-i\hbar\pa_{\!J}]
R^i(-i\hbar\pa_{\!J})Z_k[J].
\eeq
In terms of the functional $W_k[J]$ we have
\beq
\label{k5}
\qquad\delta\pa_kW_k[J]=
\pa_kS_k[\pa_{\!J}W_k-i\hbar\pa_{\!J}]
\delta\Psi_{,i}[\pa_{\!J}W_k-i\hbar\pa_{\!J}]
R^i(\pa_{\!J}W_k-i\hbar\pa_{\!J})\cdot 1.
\eeq

Finally, the gauge dependence of the flow equation is described by the equation
\beq
\label{k6}
\delta\pa_k\Gamma_k[\Phi]=
\pa_kS_k[{\hat \Phi}]
\delta\Psi_{,i}[{\hat \Phi}]
R^i({\hat \Phi})\cdot 1 .
\eeq
Therefore, at any finite value of $k$ the flow equation
depends on gauges. The same conclusion is valid for the effective average action.

But what is about the case when $k\rightarrow 0$?
Usual argument used by the FRG community to argue  gauge independence
 is related to statement that due to the property
\beq
\label{k7}
\lim_{k \rightarrow 0} \Gamma_k=\Gamma,
\eeq
where $\Gamma$ is the standard effective action constructed by the Faddeev-Popov rules,
the gauge dependence of average effective action disappears
at the fixed point.
In our opinion this property
is not sufficient to claim the gauge independence at the fixed point.
The reason to think so is the  flow equation  which includes
the differential operation with respect to the IR parameter $k$.

Indeed,
let us present the  effective average action in the form
\beq
\label{k8}
\Gamma_k=\Gamma+k H_k,
\eeq
where functional $H_k$ obeys the property
\beq
\label{k9}
\lim_{k \rightarrow 0}H_k=H_0\neq 0.
\eeq
Then we have the relations
\beq
\label{k10}
\pa_k\lim_{k \rightarrow 0} \Gamma_k=0,\quad
\lim_{k \rightarrow 0} \pa_k\Gamma_k=H_0.
\eeq
These two operation do not commute and the statement of gauge independence
at the fixed point seems groundless within the FRG approach.
Due to this reason it seems as very actual problem for the FRG community
to fulfil calculations  of the  effective average action at the fixed point
using, for example, a family of gauges with  one parameter  and
choice of two different values of the parameter.

\section{Discussions}

The gauge dependence problem
in the framework of FRG approach
has been analyzed. The standard quantization of Yang=Mills theories
within the Faddeev-Popov method
is characterized by the BRST symmetry which  governs  gauge independence
of S-matrix elements. In turn the BRST symmetry is broken in the FRG
approach with all negative consequences for physical interpretation of results.
But usual reaction of the FRG community with this respect  is that they are only
interested in the effective average action evaluated at the fixed point
where the gauge independence is  expected. One of the goals of this
investigation  was to study
the gauge dependence of the  effective average action as a solution
of the flow equation.
For the first time the equation describing the gauge dependence of flow equation
has been explicitly derived. It was found the gauge dependence of flow equation
at any finite value of the IR parameter $k$. The FRG cannot be considered as
a suitable quantization method of gauge theories given physically meaningful results.

\section*{Acknowledgments}
\noindent The work  is supported by Ministry of Science and High
Education of Russian Federation, project FEWF-2020-0003.


\begin {thebibliography}{99}
\addtolength{\itemsep}{-8pt}


\bibitem{Jac}
R. Jackiw,
{\it Functional evaluation of the effective potential},
Phys. Rev. {\bf D9} (1974) 1686.

%\bibitem{DJac}
%L. Dolan, R. Jackiw,
%{\it Gauge invariant signal for gauge symmetry breaking},
%Phys. Rev. {\bf D9} (1974) 2904.

\bibitem{Niel}
N.K. Nielsen, {\it On the gauge dependence of spontaneous symmetry
breaking in gauge theories}, Nucl. Phys.  {\bf B101} (1975) 173.


\bibitem{FP}
L.D. Faddeev, V.N.  Popov,
{\it Feynman diagrams for the Yang-Mills field},
Phys. Lett. {\bf B25} (1967) 29.


\bibitem{LT1}
P.M. Lavrov, I.V. Tyutin,
{\it On the structure of renormalization in gauge theories},
Sov. J. Nucl. Phys. {\bf 34} (1981) 156.


\bibitem{LT2}
P.M. Lavrov, I.V. Tyutin,
{\it On the generating functional for the vertex functions in Yang-Mills theories},
Sov. J. Nucl. Phys. {\bf 34} 1981) 474.

\bibitem{BV}
I.A. Batalin, G.A. Vilkovisky,
{\it Gauge algebra and quantization},
Phys. Lett.  {\bf B102}, 27 (1981).


\bibitem{BV1}
I. A. Batalin, G. A. Vilkovisky,
{\it Quantization of gauge theories with linearly
dependent generators},
Phys. Rev.  {\bf D28} (1983) 2567.%-2582.

\bibitem{VLT}
B.L. Voronov, P.M. Lavrov, I.V. Tyutin,
{\it  Canonical transformations and gauge dependence in general gauge theories},
Sov. J. Nucl. Phys. {\bf 36} (1981) 292.

\bibitem{Wet1}
C. Wetterich,
{\it Average action and the renormalization group equation},
Nucl. Phys.  {\bf B352} (1991) 529.

\bibitem{Wet2}
C. Wetterich,
{\it Exact evolution equation for the effective potential},
Phys. Lett. {\bf B301} (1993) 90.

\bibitem{Paw}
J.M. Pawlowski,
{\it Aspects of the functional renormalization group},
Ann. Phys. {\bf 322} (2007) 2831.

\bibitem{Rost}
O.J. Rosten,
{\it Fundamentals of the exact renormalization group},
Phys. Repots. {\bf 511} (2012) 177.

\bibitem{Gies}
 H.Gies,
 {\it Introduction to the functional RG
 and applications to gauge theories},
 Notes Phys. {\bf 852} (2012) 287.

\bibitem{KO}
T. Kugo, I. Ojima,
{\it Local covariant operator formalism of non-abelian gauge
theories and quark confinement problem},
Prog.Theor.Phys.Suppl. {\bf 66} (1979) 1

\bibitem{LSh}
P.M. Lavrov, I.L. Shapiro,
{\it On the functional renormalization group
approach for Yang-Mills fields},
JHEP {\bf 1306} (2013) 086.

\bibitem{L}
P.M. Lavrov,
{\it Gauge (in)dependence and background field formalism}
Phys. Lett. {\bf B791} (2018) 293.

\bibitem{LRNSh}
P.V. Lavrov, E.A. dos Reis, T. de Paula Netto, I.L. Shapiro,
{\it Gauge invariance of the background average effective action},
Eur. Phys. J. {\bf C79} (2019) 881.

\bibitem{Lav2}
P.M. Lavrov,
{\it RG and BV-formalism},
Phys. Lett. {\bf B803} (2020) 135314.

\bibitem{BLRNSh}
V.F. Barra, P.M. Lavrov, E,A, dos Reis, T. de Paula Netto, I.L. Shapiro,
{\it Functional renormalization group approach and gauge dependence in
gravity theories},
Phys. Rev. {\bf D101} (2020) 065001.

\bibitem{Lav1}
P.M. Lavrov,
{\it BRST, Ward identities, gauge dependence and FRG},
arXiv:2002.05997 [hep-th].

\bibitem{DeWitt} B.S. DeWitt,
{\it Dynamical theory of groups and fields},
(Gordon and Breach, 1965).

\bibitem{brs1}
C. Becchi, A. Rouet, R. Stora,
{\it The abelian Higgs Kibble Model,
unitarity of the $S$-operator},
Phys. Lett. {\bf B52} (1974) 344.% - 346.

\bibitem{t}
I.V. Tyutin,
{\it Gauge invariance in field theory and statistical
physics in operator formalism},
Lebedev Institute preprint  No. 39 (1975), arXiv:0812.0580 [hep-th].




\end{thebibliography}
\end{document}